\documentclass{article}
\usepackage{spconf,amsmath,epsfig}
\usepackage{diagbox}

\title{PolSF: PolSAR image datasets on San Francisco}
\name{Xu Liu, Licheng Jiao, Fang Liu
\thanks{This work was supported in part by the State Key Program of National Natural Science of China (No.61836009, No. 91438201 and No. 91438103),
the National Natural Science Foundation of China (No. 61801351)
the National Science Basic Research Plan in Shaanxi Province of China (No.2018JQ6018),
the Fund for Foreign Scholars in University Research and Teaching Programs (the 111 Project) (No. B07048), the Fundamental Research Funds for the Central Universities ( No. XJS17108) and the China Postdoctoral Fund (No. 2017M613081).(Email: xuliu361@163.com)}}

\address{Key Laboratory of Intelligent Perception and Image Understanding of Ministry of Education,\\
International Research Center for Intelligent Perception and Computation,\\
Joint International Research Laboratory of Intelligent Perception and Computation,\\
School of Artificial Intelligence, Xidian University, Xi'an, Shaanxi Province 710071, China}
\begin{document}
%
\maketitle
\begin{abstract}

Polarimetric SAR data has the characteristics of all-weather, all-time and so on, which is widely used in many fields. However, the data of annotation is relatively small, which is not conducive to our research.
In this paper, we have collected five open polarimetric SAR images, which are images of the San Francisco area. These five images come from different satellites at different times, and has great scientific research value.
We annotate the collected images at the pixel level for image classification and segmentation. For the convenience of researchers, the annotated data is open source https://github.com/liuxuvip/PolSF.
\end{abstract}
\section{Introduction}
\label{sec:intro}
With the development of sensors, studing their differences and characteristics are significative research topics \cite{Chen2017Multi}. Different sensors produce different characteristics of data and images.
That is to say, each kind of data has its own characteristics and advantages.
Making full use of these data is significative and challenging.

In the literature, polarimetric SAR image classification is a hot topic \cite{xu2019pcn, zhang2017complex, yuweipol, bihaixiapol, chensiwei, lfpol, EP2019}. polarimetric SAR image classification is also a pixel level remote sensing image interpretation task, which has the characteristics of fine recognition and need to determine the category of each pixel.

In this paper, we introduced the data and gave our labeled ground truth. To meet the needs of researchers.
First, we download the original data from the website~\cite{polsarimage}. Second, we use the ESA PolSARpro v6.0 (Biomass Edition) software~\cite{pottier2019polsarpro, polsarpro} read the original data files and get the PauliRGB images. Third, we get the high-resolution Google map of the same period, mark the image with its information by the labelme software~\cite{labelme}. Finally, 
the marked color map is remapped and encoded to get the mark file. 
In the following chapters, dataset details will be introduced.

\begin{figure}[!htb]
  \centering
\begin{minipage}[b]{0.48\linewidth}
  \centering
\centerline{\hspace{0.3cm}\epsfig{figure= 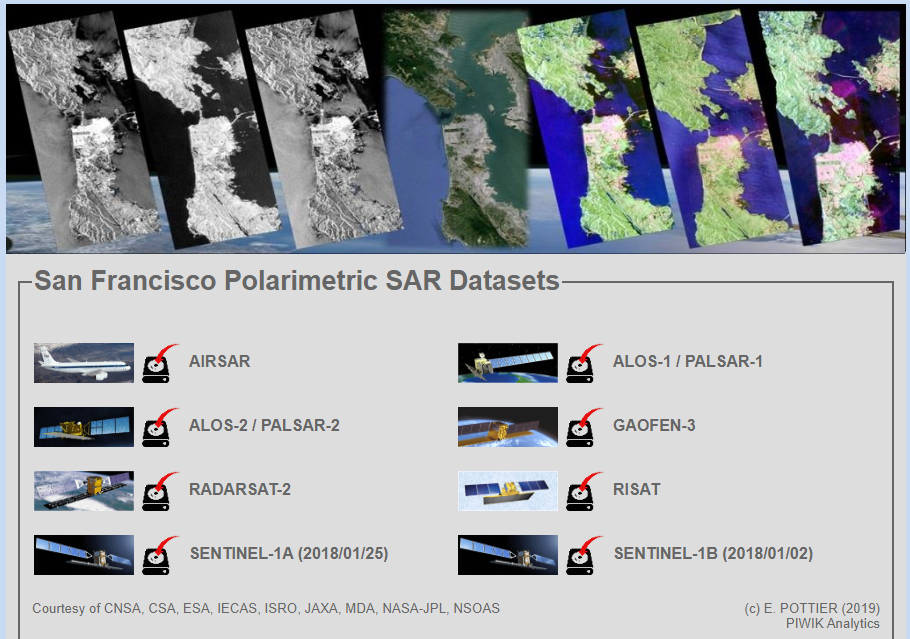,width=8.0cm}}
\end{minipage}
\caption{The polarsar data in the web.}
\label{web}
\end{figure}


\begin{figure}[!htb]
  \centering
\begin{minipage}[b]{0.48\linewidth}
  \centering
\centerline{\hspace{0.3cm}\epsfig{figure= 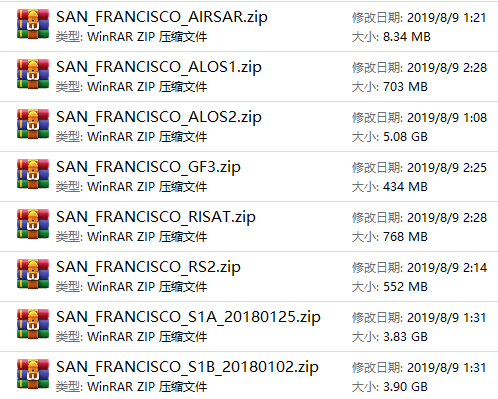,width=8.0cm}}
\end{minipage}
\caption{PolSF dataset files.}
\label{datefile}
\end{figure}

\begin{figure}[!htb]
  \centering
\begin{minipage}[b]{0.48\linewidth}
  \centering
\centerline{\hspace{0.3cm}\epsfig{figure= 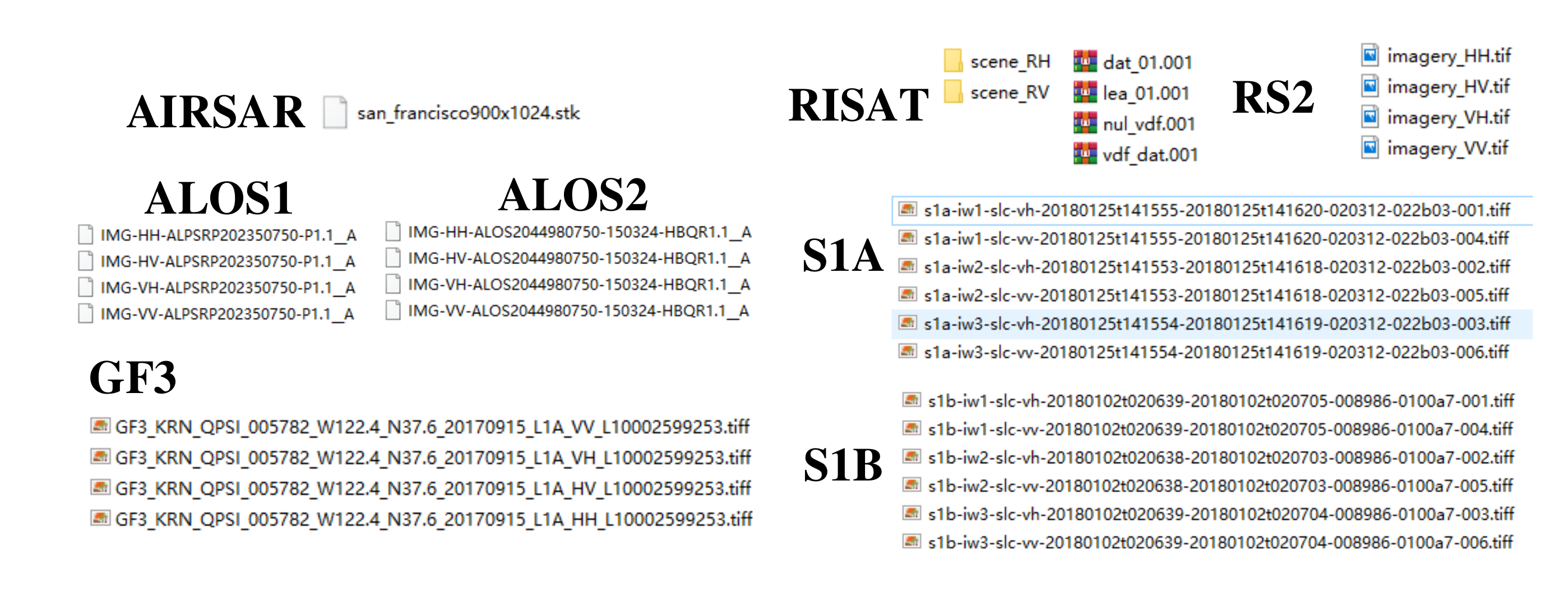,width=8.5cm}}
\end{minipage}
\caption{The data format in each satellite.}
\label{datastruct}
\end{figure}

\begin{figure*}[htb]
  \centering
\begin{minipage}[b]{0.48\linewidth}
  \centering
\centerline{\hspace{0.3cm}\epsfig{figure= 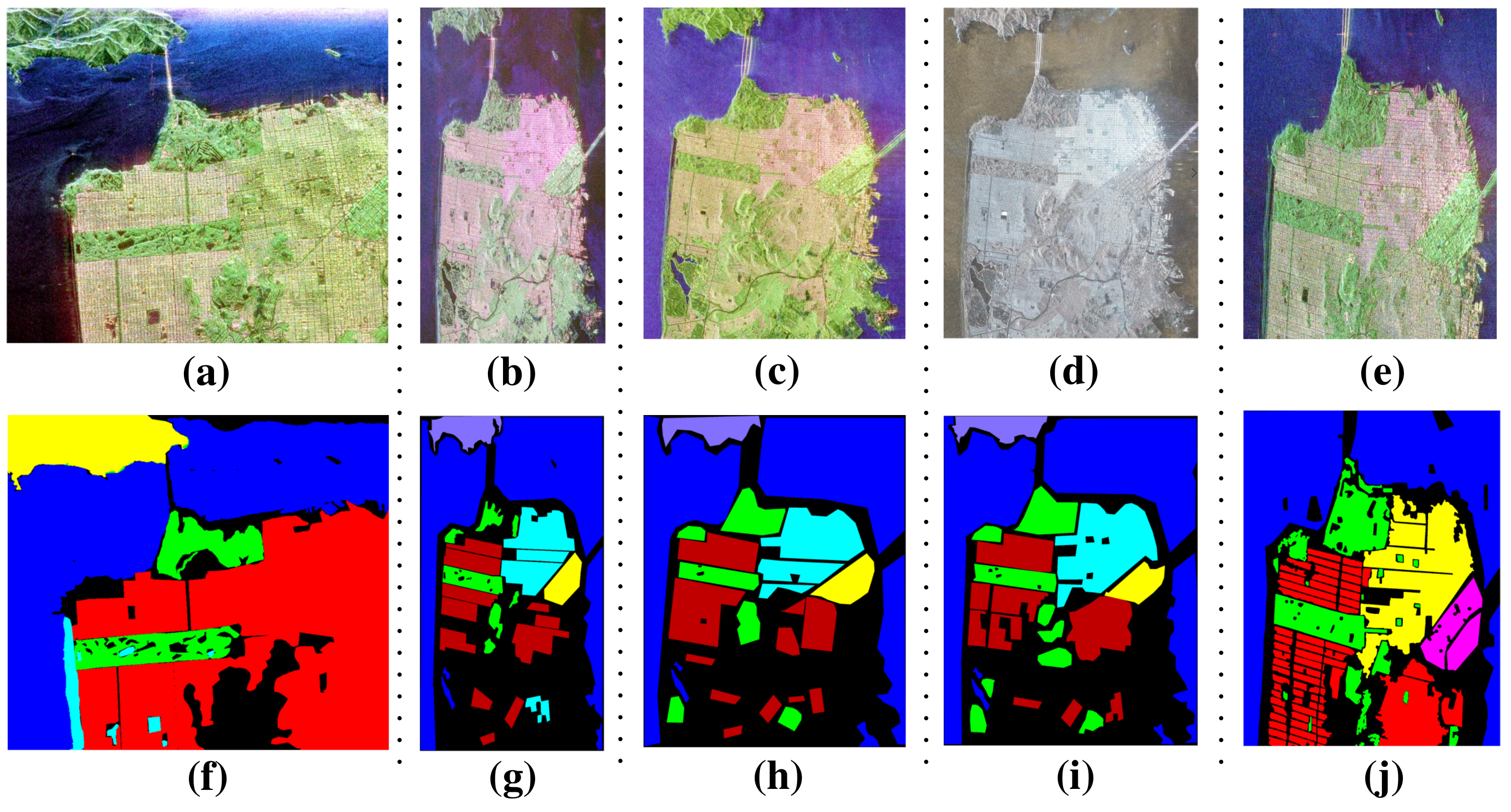,width=18.0cm}}
\end{minipage}
\caption{PolSF Dataset visualization and annotation. (a)-(e), PauliRGB image of SF-AIRSAR, SF-ALOS2, SF-GF3, SF-RISAT and SF-RS2. (f)-(j), Color ground truth of SF-AIRSAR, SF-ALOS2, SF-GF3, SF-RISAT and SF-RS2.}
\label{polsf}
\end{figure*}
\section{PolSF}
In this section, the PolSAR dataset \emph{PolSF} is presented [See Fig \ref{polsf}]. The original PolSAR data is downloaded the IETR website[See Fig \ref{polsf}]~\cite{polsarimage}. We select five the PolSAR data of five satellites for the San Francisco area, cut and mark the adjacent area. 
Fig \ref{datefile} show different raw data packages, these data are uploaded in August 9th, 2019. In addition, we can see from Fig \ref{datastruct} that the data formats of different satellite imaging are different. Table \ref{datail}  also give the details of the satelites.

\subsection{PolSF-AIRSAR}
The AIRSAR data is shown in Fig\ref{polsf}(a) and Fig\ref{polsf}(f). The imaging time is 1989. The spatial resolution is 10m. Fig\ref{polsf}(a) is a pseudo color image formed by PauliRGB decomposition. The size of original image is 1024$\times$900. Fig\ref{polsf}(b) is the ground truth marked by our team IPIU, there are five categories of objectives. i.e. 1, Montain, 2, Water, 3, Urban, 4, Vegetation, 5, Bare soil.
\subsection{PolSF-ALOS2}
The ALOS2 data is shown in Fig. \ref{polsf}(b) and Fig. \ref{polsf}(g). The imaging time is 2015. The spatial resolution is 18m. The size of original image is 8080$\times$22608, the coordinates of cut region is (x1:736, y1:2832, x2:3520, y2:7888), which is the coordinates of the upper left corner and the lower right corner. Fig. \ref{polsf}(a) is a pseudo color image formed by PauliRGB decomposition. Fig. \ref{polsf}(b) is the ground truth marked by our team IPIU, there are six categories of objectives. i.e. 1, Montain, 2, Water, 3, Vegetation, 4, High-Density Urban, 5, Low-Density Urban, 6, Developd.
\subsection{PolSF-GF3}
The GF3 data is shown in Fig. \ref{polsf}(c) and Fig. \ref{polsf}(h). The imaging time is 2018. The spatial resolution is 8m. 
The size of original image is 5829$\times$7173, the coordinates of cut region is  (1144, 3464, 3448, 6376).
Fig. \ref{polsf}(a) is a pseudo color image formed by PauliRGB decomposition. Fig. \ref{polsf}. (b) is the ground truth marked by our team IPIU, there are six categories of objectives. i.e. 1, Montain, 2, Water, 3, Vegetation, 4, High-Density Urban, 5, Low-Density Urban, 6, Developd.
\subsection{PolSF-RISAT}
For this image, the size of original image is 8719$\times$13843, the coordinates of cut region is (2486, 4257, 7414, 10648).
The RISAT data is shown in Fig. \ref{polsf}(d) and Fig. \ref{polsf}(i). The imaging time is 2016. The spatial resolution is 2.33m. Fig. \ref{polsf}(a) is a pseudo color image formed by PauliRGB decomposition. Fig. \ref{polsf}(b) is the ground truth marked by our team IPIU, there are six categories of objectives. i.e. 1, Montain, 2, Water, 3, Vegetation, 4, High-Density Urban, 5, Low-Density Urban, 6, Developd.
\subsection{PolSF-RS2}
The RS2 data is shown in Fig. \ref{polsf}(e) and Fig. \ref{polsf}(j). 
The size of original image is 2823$\times$14416, the coordinates of cut region is (7326,661,9125,2040).
The imaging time is 2008. The spatial resolution is 8m. Fig. \ref{polsf}(a) is a pseudo color image formed by PauliRGB decomposition. Fig. \ref{polsf}(b) is the ground truth, there are five categories of objectives. i.e. 1, Water, 2, Vegetation, 3, High-Density Urban, 4, Low-Density Urban, 5, Developd.
\begin{table}[htbp]
  \centering
  \caption{The details of each image.}
\begin{minipage}[b]{0.48\linewidth}
  \centering
\centerline{\hspace{0.1cm}\epsfig{figure= 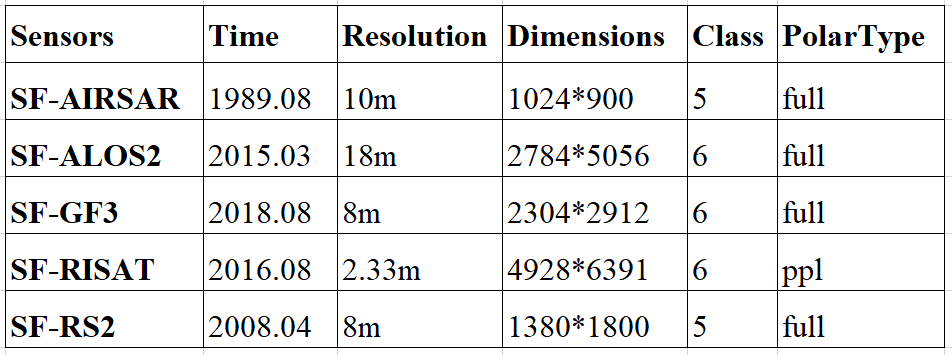,width=8.7cm}}
  \vspace{0.25cm}
\end{minipage}
\label{datail}
\end{table}
\section{CONCLUSIONS}
In this paper, we have published five image marker maps. These five images come from different satellites and have different data characteristics. There are similarities in the changes. They are suitable for further scientific research, such as single source image pixel level classification, multi-source image pixel level fusion classification, etc.
\section{Acknowledgment}
The authors would like to thank IETR provide the PolSAR data.

\bibliographystyle{IEEEbib}
\bibliography{refs}

\begin{thebibliography}{10}

\bibitem{Chen2017Multi}
Bin Chen, Bo~Huang, and Bing Xu,
\newblock ``Multi-source remotely sensed data fusion for improving land cover
  classification,''
\newblock {\em ISPRS Journal of Photogrammetry and Remote Sensing}, vol. 124,
  pp. 27--39, 2017.

\bibitem{xu2019pcn}
Xu~Liu, Licheng Jiao, Xu~Tang, Qigong Sun, and Dan Zhang,
\newblock ``Polarimetric convolutional network for polsar image
  classification,''
\newblock {\em IEEE Trans. Geosci. Remote Sens.}, vol. 57, no. 5, pp.
  3040--3054, May 2019.

\bibitem{zhang2017complex}
Zhimian Zhang, Haipeng Wang, Feng Xu, and Ya-Qiu Jin,
\newblock ``Complex-valued convolutional neural network and its application in
  polarimetric sar image classification,''
\newblock {\em IEEE Trans. Geosci. Remote Sens.}, vol. 55, no. 12, pp.
  7177--7188, 2017.

\bibitem{yuweipol}
Y.~{Guo}, L.~{Jiao}, S.~{Wang}, S.~{Wang}, F.~{Liu}, and W.~{Hua},
\newblock ``Fuzzy superpixels for polarimetric sar images classification,''
\newblock {\em IEEE Transactions on Fuzzy Systems}, vol. 26, no. 5, pp.
  2846--2860, Oct 2018.

\bibitem{bihaixiapol}
H.~{Bi}, J.~{Sun}, and Z.~{Xu},
\newblock ``Unsupervised polsar image classification using discriminative
  clustering,''
\newblock {\em IEEE Transactions on Geoscience and Remote Sensing}, vol. 55,
  no. 6, pp. 3531--3544, June 2017.

\bibitem{chensiwei}
S.~{Chen} and C.~{Tao},
\newblock ``Polsar image classification using polarimetric-feature-driven deep
  convolutional neural network,''
\newblock {\em IEEE Geoscience and Remote Sensing Letters}, vol. 15, no. 4, pp.
  627--631, April 2018.

\bibitem{lfpol}
F.~{Liu}, L.~{Jiao}, B.~{Hou}, and S.~{Yang},
\newblock ``Pol-sar image classification based on wishart dbn and local spatial
  information,''
\newblock {\em IEEE Transactions on Geoscience and Remote Sensing}, vol. 54,
  no. 6, pp. 3292--3308, June 2016.

\bibitem{EP2019}
Q.~{Yin}, W.~{Hong}, F.~{Zhang}, and E.~{Pottier},
\newblock ``Optimal combination of polarimetric features for vegetation
  classification in polsar image,''
\newblock {\em IEEE Journal of Selected Topics in Applied Earth Observations
  and Remote Sensing}, vol. 12, no. 10, pp. 3919--3931, Oct 2019.

\bibitem{polsarimage}
``{San Francisco Polarimetric SAR Datasets},''
\newblock {\em Online: https://www.ietr.fr/polsarpro-bio/san-francisco/}.

\bibitem{pottier2019polsarpro}
E~Pottier, F~Sarti, M~Fitrzyk, and Jolanda Patruno,
\newblock ``Polsarpro-biomass edition: The new esa polarimetric sar data
  processing and educational toolbox for the future esa \& third party fully
  polarimetric sar missions,''
\newblock 2019.

\bibitem{polsarpro}
``{The ESA PolSARpro v6.0 (Biomass Edition) Software},''
\newblock {\em Online: https://www.ietr.fr/polsarpro-bio/}.

\bibitem{labelme}
``{LabelMe:} the open annotation tool.,''
\newblock {\em Online: http://labelme.csail.mit.edu/Release3.0/}.

\end{thebibliography}

\end{document}